\documentclass{article}
\usepackage[paper=a4paper,left=20mm,right=30mm,top=20mm,bottom=25mm]{geometry}
\usepackage{amssymb,latexsym,amsfonts,url,rotating,dcolumn,ifthen}
\usepackage[fleqn]{amsmath}
\usepackage{pslatex}
\usepackage{epstopdf}
\usepackage{float}
\usepackage[mathscr]{eucal}
\usepackage[usenames]{color}
\usepackage[dvips]{hyperref}
\usepackage[T1]{fontenc}
\usepackage{longtable}
\usepackage{graphicx}
\usepackage{epsfig}
\usepackage{fancyhdr}
\usepackage{setspace}
\usepackage{geometry}
\usepackage{subfig}
\usepackage{multicol}
\usepackage{libertine}
\usepackage[absolute]{textpos}      % Use option showboxes to see bounding areas
\usepackage[document]{ragged2e}
\begin{document}
%\begin{frontmatter}

%%\usepackage{endfloat}
%%\usepackage[light, first, bottomafter] {draftcopy}

%\bibpunct[:]{(}{)}{,}{a}{:}{,}

\title{Threadneedle: An Experimental Tool for the Simulation and Analysis of Fractional Reserve Banking Systems.}
\author{Jacky Mallett  (jacky@ru.is)}
%\address{Menntavegur 1, Reykjavik}
%\date{}
%\end{frontmatter}
%%% BEGIN DOCUMENT
\maketitle
\section*{Abstract}
Threadneedle is a multi-agent simulation framework, based on a 
double entry book keeping implementation of the banking system's
fundamental transactions. It is designed to serve as an experimental
test bed for economic
simulations that can explore the banking system's influence on the
macro-economy under varying assumptions for its regulatory framework,
mix of financial instruments, and activities of borrowers and lenders.
Support is provided for Basel Capital and central bank reserve regulatory
frameworks, inter-bank lending and correct handling of loan defaults
within the bank accounting framework. 
\par
In this paper we provide an overview of the design of Threadneedle,
and the rational for the double entry book keeping approach used in 
its implementation.  We then provide evidence from a series of
experiments using the simulation that the macro-economic behaviour 
of the banking system is in
some cases sensitive to double entry book keeping ledger definitions, 
and in particular that loss provisions can be systemically affecting.
We also show that credit and money expansion in Basel regulated systems 
is now dominated by the Basel capital requirements, rather than 
the older central bank reserve requirements. This implies that bank 
profitability is now the main factor in providing new capital to 
support lending, meaning that lowering interest rates can act
to restrict loan supply, rather than increasing borrowing as currently
believed.  We also show that long term liquidity flows due to interest 
repayment act in favour of the bank making the loan, and do not 
provide any long term throttling effect on loan expansion and money 
expansion as has been claimed by Keynes and others.  
\section*{Introduction}
Modern banking systems sit at the centre of a complex network 
of interacting contractual and financial relationships which together comprise 
the modern monetary system. Historically analysis of this system has
presented considerable challenges, due to its complexity, its nature
as an emergent system being modified over time, and also in no small part 
to its intrinsic operational dynamics, which rely on a mixture of statistical
multiplexing and recursive feedback controls.
\par
In this paper we present a simulation framework, Threadneedle, 
which is designed to allow researchers to explore the behaviour 
of fractional reserve based banking systems and their surrounding financial
frameworks under varying assumptions about their regulatory frameworks, 
financial instruments, and patterns of borrowing and lending behaviours.  
Unlike other economic models which are based on mathematical interpretations
of the economy, or balance sheet views, Threadneedle is based on the same
double entry booking, credit and debit operations, which are used by the 
banking system's accounting processes. 
\par
We believe this approach offers a number of advantages.
It is reproducible. Banking transactions are discrete, regulatory
frameworks are defined sets of rules, and the resulting
system is computable. Given a set of double entry book keeping 
operations, and an accompanying regulatory framework, any computer simulation 
should be capable of reproducing the results of these operations with 
matching results.\footnote{Subject to known limits on decidability within 
distributed systems\cite{fischer.1985}}
In principle such simulations can also be subjected to
a standard accounting audit for verification.
It is concrete, and falsifiable.
By basing the simulation on double entry book keeping practices 
we ensure that we are reproducing the banking system at its most fundamental
level. Questions about these practices can be referred back to their
actual implementation in the banking system, and if there are 
differences between national accounting systems, or between individual
institutions, these can be incorporated into the simulation, and a determination
made as to whether they are simulation, and by implication, economically affecting.
It is scientific. Experiments can be performed on a replica
of the actual system, and causal relationships (or their absence) determined;
for example comparisons between different forms of lending within the otherwise
identical banking systems.
\par
In the rest of this paper we will discuss some of the design considerations
involved in banking system simulation, and present results from experiments
performed on simple two bank systems. Owing to some of the confusion that 
surrounds banking operations in the literature, we will start with a brief 
overview of the standard textbook description of banking, and discuss some of its errors.
We will then present some preliminary results from simulations which
examine the impact of interest rate changes in a Basel regulatory
framework, particularly with respect to liquidity flows between banks. 
Finally we will
examine the significance of these results for economic treatment
of the banking system, in particular the banking system's 
sensitivity to changes in the regulatory framework, and the 
implications this has for macro-economic models of the monetary system.
\section{Background}
Economics theories specific
to the banking system typically rest in an uneasy 
intersection between mathematical models based on
formulations derived from observational data that ignore 
any peculiar or unique role banking may play within the economy, 
and the empirical reality of periodic banking system instabilities,
significant disruption to the economy and over a century of
speculation about their precise causes.
Economic analysis of the financial system typically relies
on a set of assumptions about the behaviour of the banking system which
were developed during the 1930's. Following a remarkably periodic set
of banking crises during the 19th century, regulatory controls developed
by the British Empire after the panic of 1873 introduced a period 
of seeming stability (Laidler 2003\cite{laidler.2003}), at least within 
the British monetary system. The British
Monetary Orthodoxy as Fetter\cite{fetter.1965} later described it consequently
became a template for the regulation of banking systems world wide; it was
also embedded in a set of simplifying assumptions for economic theories 
that depicted
banking as a stable, controlled and uniform system, under the watchful 
regulatory eye of a national central bank, 
informed on how to conduct its operations by economic theory. 
\par
Consequently the banking system 
came to be regarded as a victim of financial shocks, 
rather than their cause. Bernanke in 1999\cite{bernanke.1999} for example, 
describes it as acting simply as an accelerator of endogenous developments 
in credit markets.  As such the banking system required no special analysis, 
and could be abstracted as a simple supplier of credit. 
There was no perceived need to examine the possibility that it 
could be the direct cause of economic shocks, nor to consider the idea that
bank credit carried side effects that distinguished it from other forms 
of lending.
\par
This question of whether banking is the victim of financial shocks,
their instigator, or possibly both, is a critical one for economic
theory. Treating the banking system as a benign black box, simply
supplying credit to the economy considerably simplifies the problem
of analyzing an already complex system. For this to be the case though, the
behaviour of the system must at a minimum be linear and predictable.
If non-linear and dynamic processes can be attributed to the banking
mechanisms then we must at least cast this assumption as non-proven. 
Since the banking system relies on statistical multiplexing techniques, 
and uses recursive mechanisms for control that embody positive and negative 
feedback mechanisms, some potential for intrinsic dynamism certainly exists. 
\par
Since the credit crisis of 2007 the problems with the assumption
of banking system stability have been considerably discussed, especially
with respect to current macro-economic models.
Proposals to resolve it such as Borio\cite{borio.2012} typically revolve around 
attempts to simplistically add high level observational theories to 
existing economic models, rather than re-examining the foundations of the
models themselves. These economic models rarely present 
a coherent regulatory framework, and on examination are often a mixture of 
assumptions from older work on gold standard/reserve based systems and
the post-Bretton Woods, Basel Accord capital based systems. Monetary
expansion is often attributed purely 
to government money printing (seignorage) e.g. Krugman 1998\cite{krugman.1999}),
without considering the role of changing reserve requirements, and their
effective removal in many countries.  
No distinction is made between different types of debt within
the economy, even 
though debt created by the banking system is accompanied by side effects on the 
liability deposit money supply.  A commonly expressed view, in this case
by Krugman\cite{krugman.2012} is that "the overall 
level of debt makes no difference to aggregate net worth -- one person's 
liability is another person's asset", and that it is simply the distribution of 
debt that matters (Krugman 2012\cite{krugman.2012}). This balance sheet view
ignores issues not only with the 
amount of money required to pay any given quantity of debt, but its
distribution to allow debt to be
paid. Clearly Krugman's statement cannot be true for completely arbitrary
quantities of debt, as in the limit there would be insufficient money 
across the economy
available to meet monthly principal and interest repayments. Only the
mechanisms of
fractional reserve banking guarantee matching amounts of money against
debt being created.  Other sources of debt 
such as government and corporate borrowing and loan securitization
add to the quantity of debt in the economy
without increasing the money supply.\cite{mallett.2012.1}
\par 
Perhaps most critically though, even though
specialists in banking such as Borio\cite{borio.2009} acknowledge the 
many defects in economic descriptions of the banking system's operations; 
it has become almost impossible to determine from the many claims and 
counter-claims being made about the system's behaviour, which is actually 
correct, since few attempt are made to root these claims in statements 
or descriptions that can be proved or disproved by reference to concrete 
operations derived from the banking system's book keeping 
operations.  
\par
Computer simulation, while widely used in other fields, has been
relatively under utilized in monetary economics. The Sante Fe
Stock Market Simulation is perhaps one of the best known market
simulations, but it concentrated purely on market and trading 
dynamics, and did not include any larger economic context such 
as borrowing and money supply issues\cite{lebaron.2002}. 
Lehtinen\cite{lehtinen.2007} presents an interesting 
argument which suggests that this oversight 
is a consequence of economic thought typically proceeding from top down
approaches based on
mathematical models of theories of equilibrium, 
rather than following the bottom up approach of constructing 
as realistic as possible a simulacrum of the object under study as more 
typically used in other fields.\footnote{For a recent discussion of 
the pitfalls of models in the financial and monetary context see
Pfleiderer\cite{pfleiderer.2014}, Leijonhufvud\cite{leijonhufvud.1973}, 
also provides an entertaining description of the anthropological role 
models play in the development of economic thought.}
This appears to be true even with the more recent agent based models.
Examination of their source code shows economic theories such as the 
Cobb-Douglas production function being incorporated directly into 
these models and the advice to use careful "calibration"
of their parameters to achieve desired results. 
Bianchi's\cite{Bianchi.2007} description
of the CATS model provides a typical example of these problems. 
\par
An alternate explanation for why simulation of the banking system in
particular has not been previously attempted, even though simulations of 
market trading have, may simply be the absence of reliable literature in this 
area. Our first attempts to build a banking system 
simulation (see Mallett 2011) \cite{mallett.2011}
relied on the textbook description of the banking system found in 
standard economic textbooks such as Mankiw\cite{mankiw.1997}
Burda and Wyplozs\cite{burda.2013}.
The result of this work was to demonstrate that the textbook 
description could not in fact be implemented as shown, and our
conclusion was that this description could not be taken as a reliable
model of the banking system's behaviour.  
\par
The description itself appears to originate from an explanation of the 
deposit expansion process provided in the 1931 Macmillan report to the British
Parliament\cite{macmillan.1931} shown here in table \ref{tab:expansion}, which
is believed to have been authored by Keynes.
The particularly problematic aspect of the modern description seems 
to arise from a copy and paste error between Keynes'
original description, where he restricted his example very specifically 
to "occurring in a single bank", and an expanded example where loans are 
shown being made between a series
of banks, originating from one bank and being made to a customer
at the next bank in the series.
This example breaks down immediately loan repayments are applied. 
Banks are left with insufficient liquidity
to transfer customer re-payments back up the chain after a couple
of repayment cycles owing to the leverage created between reserves and
deposits. We surmise that Keynes was aware of this issue, 
since he was careful of his
wording, but that subsequent authors apparently were not. It
also follows though that this description cannot be 
generalized as a description for banking systems that do not
consist of a single bank.
\par
There are other problems with the canonical description, which 
when considered in the context of double entry book keeping ledger definitions
cannot be excused so lightly, as they 
appear to violate the fundamental accounting equation\ref{act_eqn}: 
\begin{eqnarray}
Assets = Liabilities + Capital
\label{act_eqn}
\end{eqnarray}
\par
\begin{table}[ht]
\centering
\begin{tabular}{l|c|c|c}
Bank & Amount Deposited & Loans & Reserves \\
A    & 100              & 90    & 10       \\
B    & 90               & 81    & 9        \\
C    & 81               & 72.9  & 8.1      \\
D    & 72.9             & 65.6  & 7.29     \\
\multicolumn{4}{c}{etc.} 
\end{tabular}
\caption{Textbook Description of Deposit Expansion with a 10\% reserve requirement}
\label{tab:expansion}
\end{table}
We can deduce from Table \ref{tab:expansion} 
that the initial deposit must be a cash
deposit since reserves are being shown as withheld from it. 
Under double entry book keeping this would be defined as an asset, a bank loan 
is also an asset as is the central bank reserve account (or a reserve of 
physical cash).  The liability deposit that is in actuality being
created by the banking system when the loan is issued is not shown. 
Consequently even the original example it seems, must at best be regarded 
as incomplete. 
\par
While the inadequacies of this introductory description are being
acknowledged, most recently by the Bank of 
England\cite{mcleay.2014}, the task of replacing it has yet to be
attempted.  Documentation at this level of detail for the banking system's 
double entry book keeping practices is not easy to find - the old manual
processes that were employed to maintain double entry book keeping
ledgers and records were computerized in the 1960's, and are now
relatively remote from day to day practices.
A number of detailed descriptions of 19th century book keeping practices 
specifically written for banks do exist:
Alexander Shand's Ginko Bohi Seiho\cite{shand.1874} provides
a description of the seven fundamental operations of banking and was used 
by the Japanese Empire to convert their financial
system entirely to double entry book keeping shortly after the Meiji 
Restoration. Meelboom\cite{meelboom.1904} 
describes both the ledger and the organizational practices and procedures
of what was at that time an entirely manual operation in considerable detail
for an American bank, but does so shortly before the founding of the 
Federal Reserve and so does not include the operations of the central bank.
\par
While more recent information can be found on isolated
operations, we were unable to find any modern
documentation suitable for our purpose. Mecimore's Bank Controller's Manual
from 2005\cite{mecimore.2005} for example, contains detailed treatment of 
US regulations and account treatment but stops just above the double entry 
book keeping level of accounts.
\par
To address this issue, we have created a 
detailed description of the fundamental bank book keeping operations which are
used by the framework, modelled on Shand's approach. This has been 
reviewed by subject matter experts in bank accounting, and 
is available on line\cite{mallett.2012.2} at the arxiv.org
repository for reference. Worked examples for each 
double entry book keeping operation supported by the simulation framework
are provided in order to ensure that our understanding of these
operations is correct.  While every effort has been made to 
validate the operations described, the design of the simulation framework 
also allows any errors to be easily corrected. 
\section{Design Considerations}
\begin{figure}[H]
\begin{center}
\includegraphics[width=14cm]{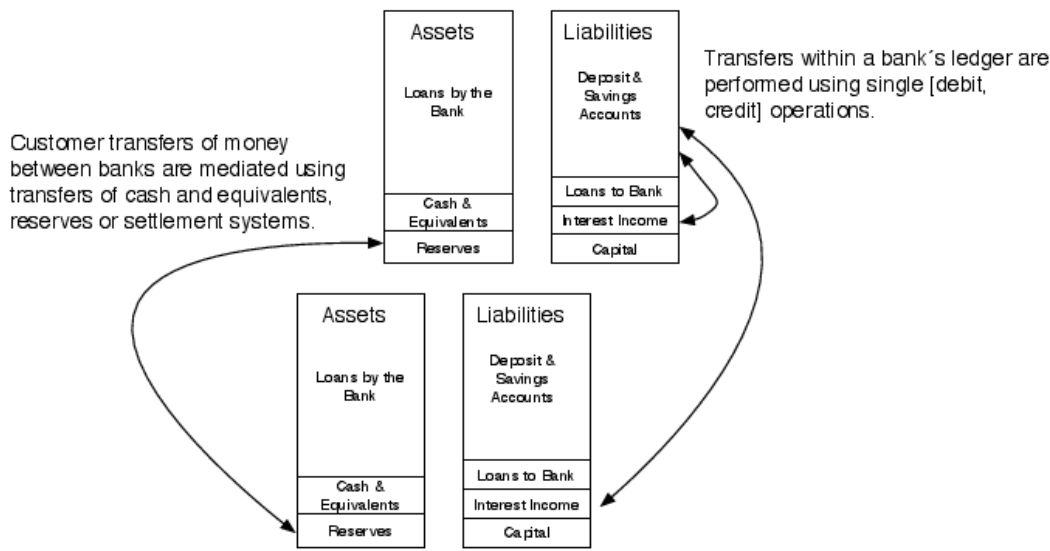}
\caption{Flows within the Banking System}
\label{fig:banking_system}
\end{center}
\end{figure}

Threadneedle presents the banking system from a ledger perspective,
directly matching the underlying books used for double entry 
book keeping operations.  From this perspective, flows within the
banking system originate from transfers between accounts, always
performed as two simultaneous operations on separate ledgers.
For example, 
Figure \ref{fig:banking_system}
shows a high level ledger view of two banks, and the monetary flows 
that accompany repayment of interest on a bank loan both within and between 
them.\footnote{In this paper we simplify bank operations to a single
ledger representing one bank. In practice banks normally operate as
a collection of branches, each of which maintains its own general ledger, and
inter-branch operations are required to move funds between branches
in similar ways to the movement of funds between banks. The impact
and variability of such operations is currently an open research
question, (See Gudjonsson 2014\cite{gudjonsson.2014}).}
All operations on bank ledgers are
performed using double entry book keeping operations, each of which consists
of a (credit, debit) tuple, which must be applied to two separate
ledgers simultaneously. One of the consequences of this arrangement
is that it effectively creates two separate forms of money, asset
cash deposited at a bank, and liability deposit accounts, which 
can be created either by cash deposits, but also by bank lending. Transfers
between monetary accounts can only be done on a like for like basis. 
It is possible to transfer directly between two liability accounts
at the same bank, and it is possible to transfer between two asset
cash accounts, by removing money from one, and re-depositing it at the other,
but it is not possible to transfer from a liability deposit account to
an asset cash account in a single operation. The two forms of money are
maintained completely separately by the underlying operations.
\par
Threadneedle is structured around this ledger view of the banking
system, with banks that individually implement the necessary double entry 
book keeping operations on a transactional basis. Loans for example are
simulated as a transaction that initially credits
the borrower's deposit account, and debits the asset loan account, and 
then a sequence of debits to the loan's capital
and credits to the bank's interest income account from the borrower's
deposit account for each payment period of the loan.
Within the simulation agents 
with distinct behaviours required by the banking system, such 
as capital purchase, are
provided and these also provide templates that can be extended to build
more sophisticated 
agents as required.  All communication between agents is performed 
via monetary transactions, that is the agents only respond to the information
provided by monetary flow within the simulation and implicit information
derived from it such as interest rates and prices. A government
agent is responsible for setting the base interest rate through
the central bank, government borrowing and taxation. These details can 
also be controlled by configuration, and changed during the simulation. The
framework also provides support for the purchase and sale of arbitrary items
directly between agents, or through market mechanisms.
\par
In order to create an environment where the results of different
regulatory frameworks on the limits of the money and loan creation 
processes inherent within banking can be studied in isolation, the
framework provides support for the banking system to be exercised
independently of a full economy.
The eventual goal of the project though
is to allow large scale simulation of all monetary transactions in a 
market based economy including an accurate representation of the banking 
system, and features have been built into the framework to support this 
eventual goal as well as the immediate requirements of
banking system simulation. 
\par
All agents in the simulation can be provided with individual
behaviours. This includes organizations such as countries and regions,
which can support different tax regimes, and can also provide structural 
associations between agents. The simulation
currently supports a single country, currency, and associated central
bank, but multiple regions can be defined within a country. Agents can be 
restricted to only operate within a single region - allowing geographical 
structure to be explored. Relationships can also exist between
agents, for example companies can employ agents, and in doing so provide
income. Inheritance is used widely, so that heterogeneous classes
of agents can also be developed, with similar functions but different
behaviours within the simulation where this cannot be adequately provided
by configuration parameters.
\par
Two base classes of bank are
currently supported: one is a non-lending bank which functions
purely as a deposit holder, and does not perform fractional reserve
lending.  This is provided for testing purposes and also allows 
experiments with constant money economies to be performed. The
other supported bank provides an implementation of the Basel regulatory 
framework, with lending controlled by a risk based weighting
of its loan book, and central bank reserve requirement. Both of
these can be configured to a range of values, or disabled.
Borrowers can be configured to request fixed period, fixed rate compound, 
simple interest or Icelandic indexed linked loans as required. 
Banks in the simulation currently operate with a single general ledger,
as branch banking operations are not included.
\subsection{Cash Handling}
Since all double entry transactions consist of a (debit, credit) tuple,
the simulation must provide a matching asset debit, 
or liability at the central bank if the deposit is within the clearing system.
While the (asset cash, liability deposit) pairing is straightforward, handling for
other accounts such as capital is less clear.
Table \ref{tab:setup} shows the 
pairings used for simulations in this paper. For commercial
banks cash from investors or depositors is initially placed in 
an asset cash account, and then transferred to the reserve account 
as required to meet regulatory requirements.
\par
\begin{table}[ht]
\centering
\begin{tabular}{p{3cm} p{3cm} p{4.5cm}}
        & Asset   & Liability \\
\cline{2-3} 
Commercial Bank   & Cash    &  Deposit   \\
                  & Cash    &  Capital   \\
\\
Central Bank      & Cash    &  Reserve account \\
                  & Cash    &  Deposit Account (Government) \\
\end{tabular}
\caption{Asset/Liability Double Entry Ledger Pairings used in Initialization}
\label{tab:setup}
\end{table}
\par
There are other ways to start a bank, the method
chosen above is derived from examination of the Basel Framework and its
requirement for bank capital as a regulatory factor, rather than those
used by earlier systems.  Miner's
1902 manual on bank book keeping\cite{miner.1902} for example, shows an 
alternate possibility where US Bonds are purchased as an asset in order 
to obtain circulating notes. This example predates the 
establishment of central banking in the USA, and also requires that 
government debt already exists. This is not the case when the simulation is 
being initialized since it introduces a circular dependency: for
government debt to exist, the government must be able to borrow money from 
existing deposits in the system. 
\subsection{Expansion from Initial Conditions}
The initial expansion of the system poses a particular problem
for simulation, since our goal is to deterministically explore the
behaviour of a mature banking system from a known set of initial 
conditions. Owing to the co-dependent relationship between loans
and deposits in the system, we view it as
impractical to try and attempt to start the system in its mature stage, 
absent detailed knowledge of the loan book of banks in the 
system.\footnote{The design of the system does not preclude this
possibility should access to this level of information become possible.}
\par
We can gain some insights into the underlying process if we examine 
what happens in the United States banking system system when a new bank 
is established.\footnote{The Federal Reserve provides a detailed overview
of the process for minority-owned institutions at its partnership 
for progress site\url{http://www.fedpartnership.gov/bank-life-cycle/start-a-bank/index.cfm}} 
The owners of the bank are required to provide capital 
in the form of an initial deposit of cash (asset) money, for which 
they receive shares (a liability) in the bank. In addition 
separate cash deposits are also required which will 
provide additional asset cash reserves. Presumably transfers
from deposit accounts at other banks will also serve. As lending then
takes place, the bank's liability deposits expand over time to the maximum
permitted by whichever regulatory framework it is operating under, subject
to its interactions with its borrowers, depositors and other banks. The 
federal reserve expects the bank to take three years 
to establish itself, and working capital has to be available for that period.
Further expansion of its loan book beyond its initial capital in a 
Basel regulated regime then requires the bank to increase its capital 
holdings, and possibly its central bank reserve holdings.
\subsection{Designing Banking Simulations}
The design of individual simulations begins with creating banks 
and populating them with a mixture of borrowers, savers and investors. 
Investors buy capital (preferential
shares) from the bank in exchange for a cash deposit, Borrowers are
programmed to request a loan each round until one is granted, and then
to repay it if they have funds to do so. Once they have repaid a loan
completely they will then attempt to borrow again. Borrowers can also 
be given a cash 
amount to deposit, and receive a matching deposit in their account if they 
do so.  Savers deposit cash, 
and receive a matching deposit in return. They can be used to provide
asset liquidity without additional side effects as desired.
\par
Banks cannot however be so populated ad hoc. In a fractional reserve
banking system lending creates a matching liability deposit, leading to the
well known expansion of the money supply (as denominated in liability deposits).
This initial expansion can distort the subsequent behaviour of the simulation,
depending on how it is distributed and the regulatory framework being
applied. 
\par
\begin{figure}[H]
\begin{center}
\includegraphics[width=10cm]{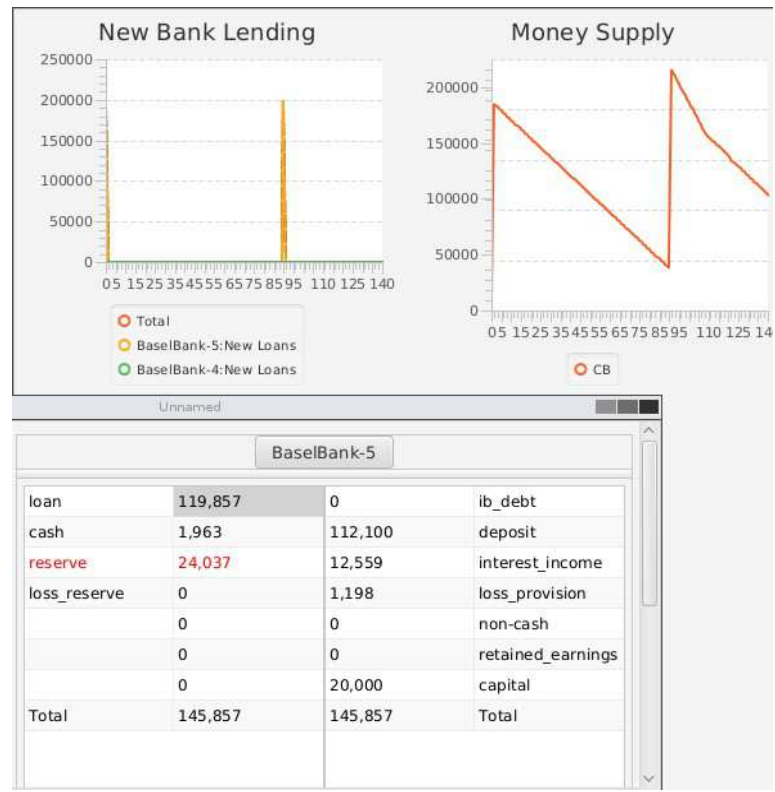}
\caption{Unevenly distributed lending in a strictly reserve regulated system}
\label{fig:reservecycle}
\end{center}
\end{figure}

If we take a slightly pathological case as an example. 
In a strict reserve regulated simulation, if the bank allows a 
single borrower to borrow the maximum amount
allowed, then the entire loan book will consist of a single loan. Until
this loan has been sufficiently repaid to free up the necessary loan 
capacity, no new loans can be made,
and the money supply will be seen to contract.
Figure \ref{fig:reservecycle} shows a single bank simulation, with 
borrowers configured to request 120 step loans of 200,000. 
Cyclic behaviour of this nature is typical of simulations of strictly reserve 
regulated
systems with insufficient or poorly distributed lending patterns,
owing to the feedback in the regulatory control between lending limits and
deposits.
Since we can deduce from modern national monetary statistics that fractional 
banking systems are typically in a state of continuous expansion, and
as far as we are aware no banking system has ever been regulated solely
on an absolutely fixed reserve of cash holdings, we can reasonably attribute this as a 
simulation artifact.
\par
In order to force borrowing to be more evenly distributed, 
borrowers can be assigned a loan window which restricts 
the steps in which they can request loans. This in conjunction with
appropriately sized loans allows more realistic simulations to be 
created.
\par
Additional considerations surround asset cash availability. In a 
reserve regulated system sufficient asset cash must be present in proportion
to the size of the loans being made. In both capital and reserve systems
there must also be sufficient cash to meet
any liquidity requirements arising from inter-bank transfers. 
In a capital
regulated system sufficient capital must be present, again in proportion
to the size of loans being requested. To create an 
even loan distribution which maximize lending, and its associated deposit
creation over time, against the regulatory framework in operation, the
following guidelines are suggested for simulations designed to saturate
the banking system's loan capacity.
\begin{align*}
\text{Loan Window}  &= L \\
\text{No. of Borrowers} &>= 5*L \\
\text{Asset Cash} &>= L * D * R  \\
                  &>= Capital * C
\end{align*}
where:
\begin{itemize}
\setlength\itemsep{0em}
\item[L=] Loan period in steps
\item[D=] Average loan amount
\item[R=] Central Bank Reserve requirement as a percentage.
\item[C=] Capital Reserve requirement as a percentage.
\end{itemize}
Under these guidelines with 
each borrower only taking one loan at a time, for 10 year 
loans (120 steps) of 10,000 monetary units each, a minimum configuration 
for a 10\% reserve requirement system would be 600 borrowers. Borrowers
should also receive a minimum deposit that allows them to make their first
loan repayment.
\par
Capital purchases act to provide the capital
requirement, so simulations testing Basel banking systems will in 
practice need to distribute the initial cash created for the system
across investors and borrowers. Savers can be used to provide cash
deposits without any further interaction with the banking system 
if desired, since the base base saver class does not receive interest
on their account. It is expected that the saver class will
be extended to allow interest bearing accounts, but this creates
accompanying flow considerations as discussed below.
\subsection{Flow Considerations in Simulation}
Simulations only run continuously as long as their monetary flows 
can be satisfied. Debtors for example, must receive some form of income source 
so that they can meet the repayments on their loans, adequate loss 
provisions must be available against which loans that default
can be written off.  
For example, if treasuries are used in the simulation, but
the government has no form of revenue, such as taxes, the simulation
will quickly halt as monetary flow breaks down due to the government
having no income to meet interest and capital payments on its 
treasuries. In simulations where flow breaks down the cause may lie
in economic fundamentals which reflect the actual
behaviour of the monetary system, or in artifacts of the simulation's 
design, and neither source should be ignored when exploring the reasons
for this phenomena. 
\par
For simulations that isolate the banking system - such as the ones
in this paper - it is necessary to provide a flow of money to a bank's 
borrowers in order that they can meet their loan obligations. To achieve
this borrowers
can be configured to be employed by a bank which then uses
its interest income to pay them  an amount that allows 
them to meet their capital and interest repayments for that round. 
Provided that the bank has sufficient interest income to meet its
borrower's requirements, this provides a continuous flow of money
between borrowers and banks which can be used to satisfy the 
constraints being set by the banks lending behaviour, and exercise
the regulatory framework.
\par
\begin{figure}[H]
\begin{center}
\includegraphics[width=14cm]{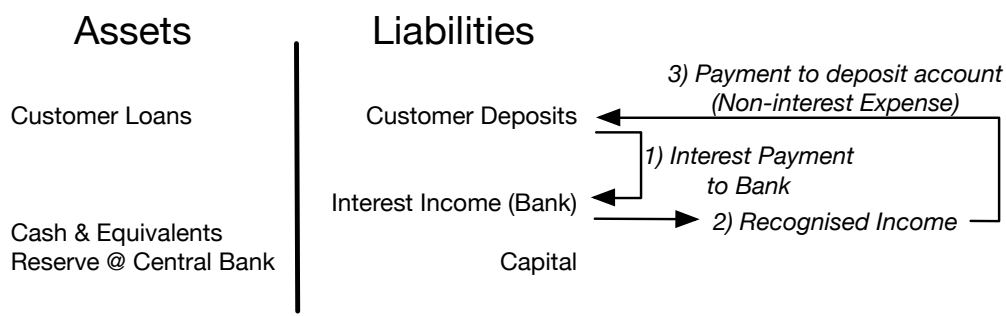}
\caption{Bank's Interest income flow through its Ledgers}
\label{fig:ii_flow}       
\end{center}
\end{figure}

To set this within an economic context, Figure \ref{fig:ii_flow} shows an 
abbreviated example of the flow of 
money within a single bank, which results from interest payments on loans
made by the bank, with the assumption that recipients of non-interest expense 
payments maintain their deposit account at the same bank as the 
account originating the interest payment. Payment of interest on bank
loans when the depositor's account is at the same bank as that making
the loan is simply:
\par
\begin{center}
(debit\ customer\ account,\ credit\ bank's\ interest\ income\ account) 
\end{center}
\par
since both accounts are classed as liabilities. When an interest (or
capital) payment is made from an account that is not at the same
bank, a more complex sequence of operations occurs, specifically:
\par
\begin{table}[ht]
\centering
\begin{tabular}{p{3cm} p{7cm}}
@ Originating Bank:& (credit cash, debit customer account) \\
@ Receiving Bank:  & (debit cash, credit bank's interest income account) \\
\end{tabular}
\end{table}
with the accompanying asset transfers occurring either through central
bank accounts, or dedicated clearing mechanisms. The availability
of asset forms of money to support interbank operations is consequently
a critical aspect of banking operations, but not necessarily an economic
one, leading to the critical distinction in banking interventions between
insolvency and illiquidity.
\par
Typically a bank
will receive interest payments into its interest income account, from
which expenses such as loan defaults must be deducted before it can 
be recognized as income and paid out as more general expenses such 
as employee salaries, dividend payments to shareholders,
and purchases of additional capital. A stochastic 
loan default rate can also be applied, and this can be used to create 
simulations where the sensitivity of the system to default risk 
can be explored.
\par
Within the larger economy, a series of payments can consequently be 
envisioned between 
the recipient of the expense payment by the bank, and the ultimate
debtor who makes the interest payment on their loan.  In the simulations 
presented here we effectively short circuit this series to a direct
relationship between a borrower and a bank, although not necessarily 
the same bank as the loan is made from.  Only in the case that the bank 
has insufficient income to make payments will borrowers in the simulation
be unable to 
pay their loans. Configuring borrowers to receive salaries from a different
bank to which they are borrowing from allows the interbank transfer
mechanisms to be exercised, and can also be used to create deliberately
unbalanced interbank flows in order to explore the
effects of interest changes on the interbank lending 
mechanisms under controlled conditions, as we will see below. 
\par
This is clearly an artificial arrangement, but it is designed to
allow the banking mechanisms to be exercised independently of the larger
economy. It also creates a best case for the banking system, since all
interest income received by the bank will be looped into loan repayments
as necessary. If regulatory mechanisms are unstable under this arrangement, 
then we can justifiably point to
the mechanisms as a cause of problems rather than the nebulously 
defined catch all of human economic behaviour. 
\subsection{Central Banking}
The Central Bank side of the system is becoming increasingly complex, especially
with the fallout from corrective measures taken during the 2007 credit crisis.
For example, the Federal Reserve's 2011 
Annual Report shows the majority of its assets as a mixture of 
government treasuries and mortgage backed securities, the latter having been 
acquired through
the Troubled Asset Relief Program (TARP).\footnote{Federal Reserve Annual 
Report 2011, p327 \url{http://www.federalreserve.gov/publications/annual-report/files/2011-annual-report.pdf}}. 
The rather less detailed balance sheet of the Bank of England
simply lists 'Other loans and advances' as the majority 
asset\footnote{Bank of England Annual Report 2011, 
p50, \url{http://www.bankofengland.co.uk/publications/Documents/annualreport/2011/2011full.pdf}} while
the European Central Bank has a mixture of gold, loans to Euro credit
institutions, and securities comprising the majority of its 
holdings\footnote{European Central Bank, Annual Report 2011 p200}. It
is not known to what extent these differences are systemically affecting.
\par
Central 
bank holdings of financial instruments that provide income, for example
treasuries, also create flows within the system that need to be balanced,
which we assume was at least part of the reason for the Federal Reserve 
to initiate interest payments on reserve holdings to its banks in the 
wake of the TARP program.
Where possible these issues have been delegated to the user and the design
of the simulation being constructed, with the framework providing as 
simple an initialization state as practical. For the time being, 
the central bank's holdings have been simplified to cash.
For the specific simulations discussed in this paper we have adopted the 
convention shown in Table \ref{tab:setup} where the central bank 
initially holds a cash asset against its liability (reserve)
accounts, and does not have a capital account. Since the main
activity of the central bank in the simulation is to support the clearing 
system through transfers between banks using their reserve accounts, we
do not believe this to be systemically affecting for the purpose
of these simulations.  The government's bank 
deposit is also held at the central bank, which appears to be the usual case 
for the banking systems we have examined. This it should be noted,
may be systemically effecting as it then interacts with the central 
bank's balance sheet, and is a topic that deserves more 
investigation.  Detailed exploration of the 
issues on the central bank side of the system is a subject for future 
research.
\subsection{Implementation}
The framework is written in Java, and can be run on Windows, Linux or MacOS.
All monetary transactions within the framework are performed using full double 
entry book keeping and a records of all transactions are available through
the simulation, and can be audited as
required. The asset, liability and capital classification of individual bank 
ledgers is configurable, which allows experiments to be performed
with different ledger classifications if required.
A set of base objects such as the Borrower
class described above are included, which
provide support for simple simulations, and can be extended by users
with programming experience to provide customized behaviours as required.
The framework currently provides the following:
\begin{table}[ht]
\centering
\begin{tabular}{p{2cm} p{12cm}}
Class       & Description \\
\hline
Govt             & Base class for governments. Flat rate tax, and government treasuries.  \\
Bank             & Provides a non-fractional reserve deposit holding bank which does not perform lending \\
BaselBank        & Provides a bank implementing Basel Regulation which can be
enabled or disabled as required.\\
Investor         & Buys bank shares (capital) and receives interest. \\
Borrower         & Borrows from a bank and receives salary. Can take out a loan. \\
Saver            & Bank deposit holder. Can be used to simply provide deposit money without side effects \\
Loan             & Fixed term, compound interest rate loan \\
Simple           & Fixed term, simple interest rate loan \\
IndexedLoan      & Icelandic Indexed linked loan \\
\end{tabular}
\label{tab:classes}
\caption{Economic actors currently provided by framework.}
\end{table}
\subsection{Configuration}
Simulations can be defined using a drag and drop interface, or
from a JSON formatted text configuration file.
Data from simulations is displayed on graphs during runtime and can 
also be exported for separate analysis.  The framework provides
a command line interface through which parameters can be examined
and changed during the simulation, and 
a batch mode. Batch mode also supports a simple programmable
interface that allows simulation parameters to be modified between
runs. Although only single country simulations are supported at 
present, countries are specified in the configuration in order to allow
eventual support for larger international simulations, either of countries
with different currencies, or multi-national currency unions.
\section{Results}
In this section we present the results from a set of
simple experimental simulations, designed to explore
some of the questions surrounding the
banking system's response to different forms of regulatory control
and macro-economic intervention such as interest rates.
In these simplified prototypical banking systems,
interest rates are set for the 
entire system, with no difference between the interbank lending rate and 
the rate for customer loans. Fixed rate loans are used, where the interest rate
applies for the entire period of the loan, i.e. the affects of interest
rate changes only take effect when new loans are issued. Long period fixed 
rate loans are typical of US bank lending, but not generally for European
banking systems. We would stress that we are
interested in this paper in the mechanical response of the banking system
at its limits to changes, rather than the results of these responses on the
economy. Ten year (120 repayment periods) loans are used
in these examples for illustrative purposes.
\subsection{Central Bank Reserve Regulation}
\begin{figure}[H]
\begin{center}
\includegraphics[width=14cm]{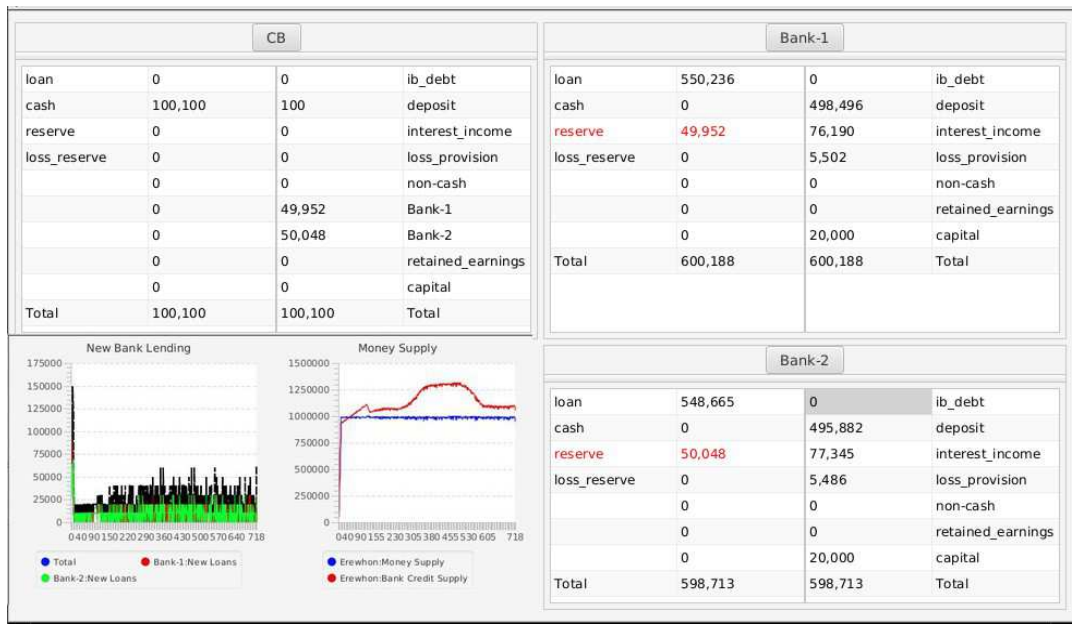}
\caption{Reserve controlled lending with changing base interest rates over time.}
\label{fig:reservelending}
\end{center}
\end{figure}

% Configuration file eea_fig4.json
Figure \ref{fig:reservelending}
shows a simple
central bank reserve regulated banking system with two banks. Each bank
has 600 borrowers requesting loans of 10,000 each, 10 year duration,
with a base rate that is initially 2\%.\footnote{Simulation 
configuration file reference: eea\_fig4.json}
The system's loan supply is saturated, as can be seen from the 
highlighted reserve constraint (show in red), and after some
perturbation as a result from the 
expansion from initial conditions (see above)  the money supply is stable. 
All lending is confined to depositors at
the loan originating bank, so there is no interbank activity.
Reserve regulation in this simulation is based on the textbook calculation
of the difference between reserves and customer deposits:
\begin{equation}
Loan\:Limit = \sum reserve\:ledger / R - \sum deposit\:ledger
\end{equation}
where R is the central bank reserve percentage, in this case 10\% providing
a theoretical (liability) money multiplier of 10. Capital controls are not
enabled.
New loans are granted when the bank's loan capacity  is greater than the capital
value of the loan. As a consequence there is some variation of the rate
of lending over time as shown in the graph of new bank lending.
With reference to the reservations previously noted, this simulation
is as close as it is possible to come to the standard textbook 
description of the banking system. In this simulation, the central bank
base rate begins at 2\%, is increased to 5\% at step 240, and returned to 
2\% again at step 480, resulting in the credit supply behaviour shown
\par
The simulation's calculation of the
reserve limit is based solely on the deposit ledger, and this in turn is
effected both by the size of ledgers derived from deposit accounts:
loss provisions and interest income.
Consequently the actual amount of lending varies directly with interest
rates, since increases in interest rates in this simulation cause an 
increase in the size of the interest income ledger which is not counted
as a deposit. This behaviour in the simulation is somewhat artificial
since typically a bank's interest income is not directly coupled to interest
rates, as changes also occur to the amount of income being paid out to
savers as an expense. However, it does
demonstrate the sensitivity
within reserve constrained banking system to ledger classifications, since this
effect will also apply to changes in loss provisions. Potentially similar
effects could occur in reserve based systems due to banks reserving against
foreseen losses or periodic dividend payments.
\par
Today banks are required to make loss provisions at the time the loan is 
made.\footnote{Full book keeping for loss provisions involves a 
contra-asset loss reserve account, the version shown here is somewhat simplified.}
The amounts held are partially regulated and partially under each
individual bank's control. Until the mid-1970's in the USA for example,
favourable tax law resulted in higher loss provisions, which were then
reduced when the law was changed. This experiment demonstrates that changes to 
loss provision requirements in conjunction with reserve regulation 
can potentially alter credit provision, depending on how they 
are classified within the system and the regulatory framework. According to
Balla\cite{balla.2012} they are accounted as a contra-asset on the
balance sheet, but since they are deducted from expenses on the income 
statement, in practice they originate from liability deposits.
Frait\cite{frait.2013} and others have suggested that loss provisioning may be
pro-cyclical within a Basel framework
however it follows from this result that this question
cannot be answered simply by looking at levels of loss provisioning, 
since the regulatory framework also plays interacts with the system's response to
these levels.
\par
\begin{figure}[H]
\begin{center}
\includegraphics[width=14cm]{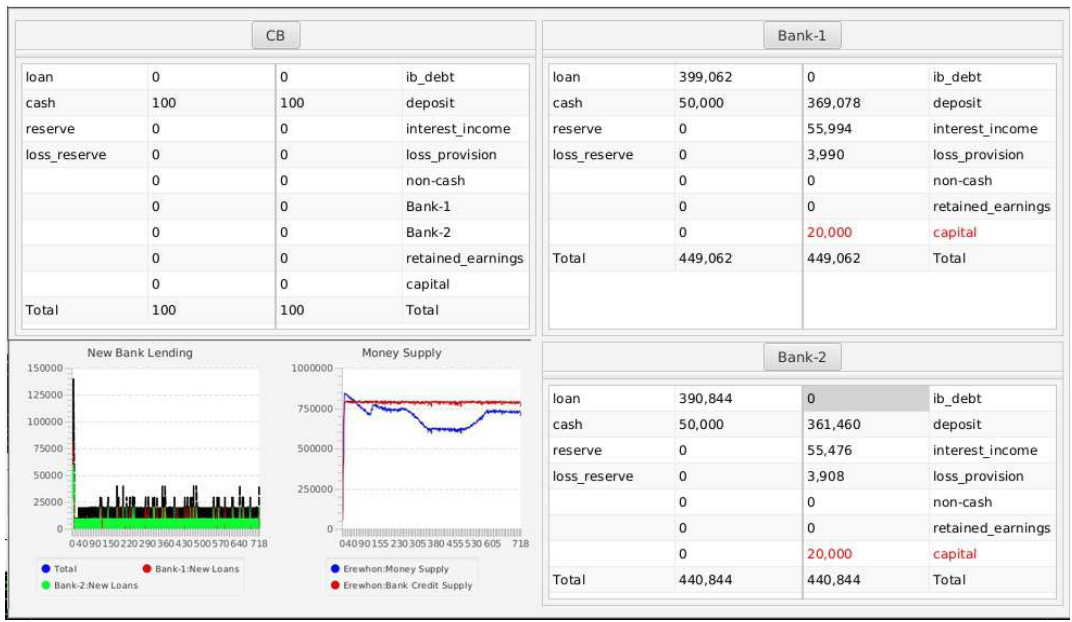}
\caption{Capital controlled lending with changing base interest rates over time.}
\label{fig:capital_lending}
\end{center}
\end{figure}

This can be seen in Figure \ref{fig:capital_lending} which in contrast shows the 
identical simulation with capital controls enabled, and reserve controls disabled. 
The capital control is a 50\% Basel risk weighting calculation applied to
all loans.
In this experiment we see no change in lending as a result of the interest rate changes,
since lending is regulated by capital holdings, and loss provisions are being treated
as a liability. Instead, the money supply decreases as a result
of the increase in interest income. This is, as discussed above, a simulation artifact,
the sum of the deposit and interest income ledgers is unchanged.
It does show that the main result of interest rate changes in a Basel regulated
system - leaving aside affects on loan demand - is to modify the distribution of monetary 
flows between savers and debtors.
\par
Treated as a liability, loss provisions do not directly interact with Basel
regulation of the loan supply, unless by reducing profits, they prevent
the bank from expanding its capital base. However, some loss provisions can 
be included in Tier 2 capital and this does have the potential to interact
with loan regulation.
\par
Also of interest is the counter-intuitive response under reserve regulation
to the increase in
interest rates, with higher interest rates leading to an increase in 
lending under.  This is again due to ledger 
classification, and the associated increase in interest income with higher 
interest rates, since the liability money in that ledger is then not
counted toward the  central bank reserve requirement. 
The simulation's mechanics represent the maximum possible size 
of the response: the actual size of the effect would be dependent on how 
much and how long interest income was retained within the banking system 
before being recognized. The possibility of individual banks manipulating
this dependency would also appear to exist.
\par
Although reserve regulation is successful in this set of artificial 
circumstances, this should not be taken as indicative of its 
actual effectiveness. This experiment did not include inter-bank 
lending, and it has been known since the 1920's that the 
regulatory role of central bank reserves could be partially circumvented by
both interbank lending and re-discounting. Keynes discusses
this as one of the results of divergent practices between the US
Federal Reserve Banks and the Bank of England in 1929\cite{keynes.1929}.
It is hard to know the full impact of these problems on the
banking systems of their time, without a complete description of their
regulatory framework, including their capital relationships, and
ledger classifications.\footnote{Under the 1844 Banking Act, London 
Banks were required to publish their holdings weekly. They developed
the practice of doing this on different days of the week, which allowed
inter-bank lending to be used to mask reserve discrepancies if necessary.}
\par
The most significant result from these simulations is the evidence that
the banking system and by extension, the macro-economy within which it 
is operating, can be affected by accounting definitions operating at the 
lowest levels of its implementation. This not only indicates a system that has
a considerably greater degree of sensitivity to the minutiae of its regulatory 
and accounting frameworks than may have previously been assumed, but also
suggests that national macro-economic variations may be in part due to seemingly
inconsequential differences in bank regulation.
\subsection{The effects of lending on bank asset liquidity}
%Configuration file: eea_fig6.json      
Modern banking systems vary considerably in their use of central bank
reserve requirements, and have generally moved away from using them
in a regulatory role. No banking system appears to apply them to bank
deposits in their entirety: at one extreme the Bank of England 
has no formal reserve requirements but has introduced liquidity requirements 
which are similar if not identical in their
effects.  The Euro banking system applies a 2\% reserve requirement to all deposit accounts, 
excepting time deposits of greater than 2 years duration, and while the US applies
a 10\% reserve requirement to net transaction accounts, in practice
there is considerable scope for account reclassification which allows
its banks significant latitude over the size of their reserve accounts.
Highly liquid forms of debt instrument such
as treasuries can also be used to meet some of this requirement.
\par
The expansion from initial
conditions show in all simulations here demonstrates that
in the absence of some form of regulation the banking system will 
rapidly expand its lending and the associated liability money supply.
If most modern banking systems do not have strict reserve requirements,
then what, besides borrower demand, is limiting the expansion of modern
banking systems?
\par
A claim made in the Macmillan Report, economic textbooks, and repeated recently
by McLeay 
et. al.\cite{mcleay.2014}\footnote{Section(i) Limits on how much banks 
can lend p5} is that liquidity considerations play a role in throttling
lending and money expansion. Specifically 
because funds being lent will be wholly or partly transferred
to other banks in the banking system, the bank must restrict its lending
to prevent it losing asset money. In the period of the Macmillan report of
course such an affect would have interacted directly with the 
formal reserve requirements applied at the time, acting as an immediate
restriction on further lending.
\par
This argument however overlooks the long term flows of money associated with
loan \emph{repayment}, and in particular interest payments. 
The total
amount of money transferred to the loan originating bank just as interest
repayment on a 25 year loan at 6.5\% is approximately equal to the original 
capital. Banks originating loans can consequently expect a net inflow
of asset money over time, even if there is a short term outflow
at the time the loan is made.
The long term and dominating effect on liquidity is consequently
the exact opposite to that claimed by Keynes et. al.
\par
\begin{figure}[H]
\begin{center}
\includegraphics[width=14cm]{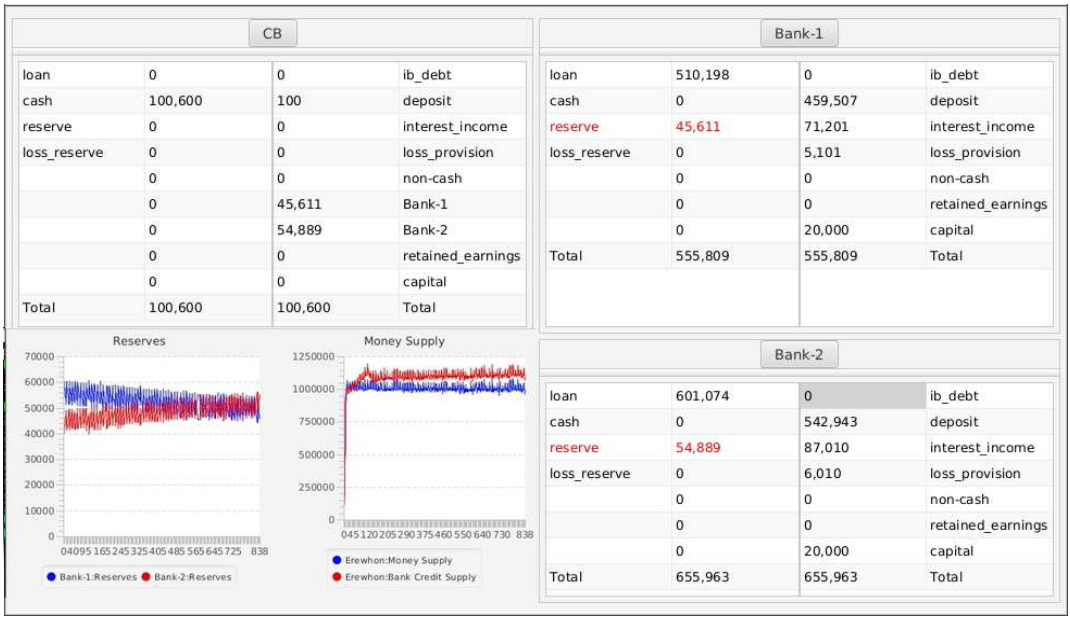}
\caption{Liquidity effects of loan related asset flows.}
\label{fig:single_loan}
\end{center}
\end{figure}

Figure \ref{fig:single_loan} shows the same simulation used in the previous
two experiments, with a single addition, a borrower with an account at Bank 1,
receiving salary payments from Bank 1, but taking out a loan of 10,000 (\~2\% of the
bank's total loan book) from Bank 2.
As shown over time, there is a steady transfer of reserves from Bank 1 to Bank
2, as interest and capital repayments are made on the loan.
\par
\par
The size of this transfer is directly proportional to
the interest rate on the loan, so as interest rates increase, the accrual
of assets at the originating bank will also increase.  This mechanism
points to an explanation for the observed effect reported by 
Ennis\cite{ennis.2001} for the USA, Benito\cite{benito.2008} for
the Spanish system, and Wilson\cite{wilson.2000} for the four largest 
European systems, namely that banking systems tend to consolidate over long 
periods of time, i.e. systems
with large numbers of small banks progressively become systems with small
numbers of large banks, unless they are specifically regulated to prevent this. 
If we simply assume a randomly distributed number of loans
across the system being paid from non-originating banks, 
we can see that probabilistically large banks will tend to cannibalize small
banks of their asset money over time. 
\par
In practice, banks monitor their liquidity very carefully, and many require
that their borrowers maintain accounts at their bank, presumably in part
because of the initial liquidity issues lending can cause.
\footnote{In Iceland, Arion Banki charges a higher interest rate to other banks' customers,
while the new MP Banki places no such restriction on its lending.}
However, banks have little or no control over
their customers activities over time. For example, a bank that deliberately
restricted lending to customers employed by a company that was also an 
account holder, would still not be able to guard against customers who
changed employers. 
Clearly a bank that was aware of this effect could potentially exploit it
competitively, but the timescale for the strategy to be 
effective would be several years. 
\par
Since increases in interest rates accelerate this mechanism, we 
hypothesize that it can be a cause of stress on the interbank lending
mechanisms, and that increasing interest rates would put additional pressure
on banks that were experiencing net asset outflows possibly leading
to bank failure. The movement within the banking system of interest
and capital flows due to securitized lending could also be expected
to provide further stress on liquidity provisions due to these factors.
Clearly though, we must look elsewhere for restrictions on the actual
amount of lending being performed by the banking system.
\par
Taken in conjunction with the result shown previously that interest
rates do not directly affect the credit supply of the system, this 
result also allows us to determine the Nyquist limit for the banking system. The
Nyquist limit is an important result from signal processing, which 
places a lower bound of twice the frequency under study, for determining 
the minimum sampling period for a time
based system which avoids the detection of spurious signals due to aliasing. 
\par
Since there appears to be no impact on supply from interest rate changes, with
the assumption that most banking systems are supply constrained, the main influence within
the system on the money and credit supply seem to stem from the creation and destruction
of money within the system which occurs as loans are made and repaid. This
implies that the average loan period in conjunction with the regulatory
limits on lending dominates in 
determining the response of the system to its regulatory framework. 
With the majority of loans in modern systems being of 20-30 year duration, 
this implies that macro-economic analysis based on monetary data must span a 
minimum of 40-60 years to avoid any spurious results due to aliasing. 
This is significantly longer than the mean time between
significant regulatory changes intended to alter
the behaviour of the system, suggesting that observational data may be
of limited benefit in assessing the long term economic consequences of any
given regulatory and accounting framework for this system.
\par
\subsection{Basel Capital Regulation}
The introduction of the Basel Accords\cite{basel.1988} 
from 1988 onwards imposed a new regulatory control on banking systems, 
one which attempted to regulate lending according to the default
risk of particular categories of loans.\footnote{See Alfriend\cite{alfriend.1988} for
a review of the shift in regulation toward capital controls from the 1940's.}
In contrast to the central bank reserve 
requirement which established a leveraged ratio between money represented as 
liability deposits, to asset deposits of 
cash\footnote{Central bank reserves are an 
asset from the perspective of the bank which owns them, and a liability
on the central bank side}; the Basel requirements establish a leveraged
ratio between a bank's capital (a liability), and its loans (an asset) 
according to a risk weighted multiplier based on the type of loan being made.
\par
There is no indication that the Basel Accords were directly intended to limit
monetary expansion, they were explicitly designed to ensure that banks
maintained sufficient capital to provide protection against insolvency 
by ensuring that banks
would have sufficient capital to handle loan defaults, in the event
that their loss provisions and profits were inadequate for this purpose.
There are no limits set on the total capital expansion by the entire banking 
system for example: 
individual banks are simply responsible for obtaining sufficient capital 
to cover their lending books with respect to the risk weighted lending 
requirements of their loan book.
\par
\begin{figure}[H]
\begin{center}
\includegraphics[width=12cm]{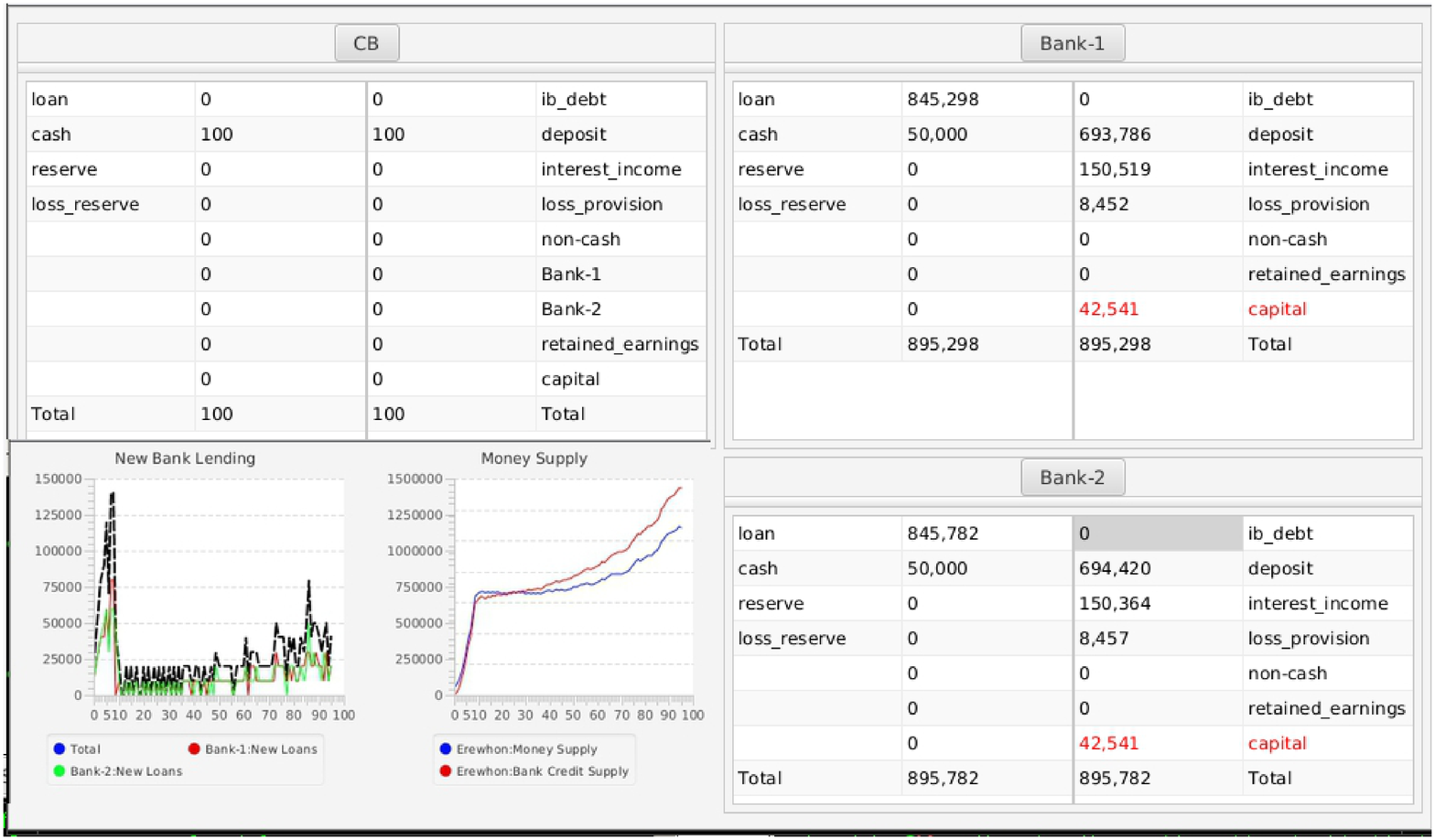}
\caption{Capital Reserve increases allowing Money and Credit expansion.}
\label{fig:capital_limits}
\end{center}
\end{figure}

This not withstanding, under simulation it is clear that the Basel capital 
requirements do exert
at least a throttle on lending and monetary expansion within the 
system, as shown in Figure \ref{fig:capital_limits}. This simulation
uses an identical configuration to the previous examples with reserve
controls disabled, and capital controls enabled. The base interest
rate is set to 5\% in order to ensure that banks have sufficient
interest income to pay dividends to their investors. In contrast to the 
previous example where no increase in capital was allowed, Investors
are paid a dividend of 5\% on their share holdings if sufficient income
is available, and can use this to purchase new capital every 12 steps. As
a consequence the credit and money supplies expand, in this instance
approximately doubling over 10 years of simulation. This also
suggests that the capital controls could be used for direct regulation
of money and credit expansion if desired, which may be worth of
further investigation, since in conjunction with risk weighting
it would also seem possible to influence sectoral lending.
\par
%
% Add calculation of basel limit
% comment on change of 10 to 5
% show that it applies
% 
% include Icelandic loans as example of financial instruments changing
% system. 
\par
Part of the answer then, to what is regulating
monetary expansion in modern banking systems, appears to be the Basel
capital requirement. While
increases to the capital on which the requirement is calculated
are not explicitly limited, they are also not completely arbitrary. Increases
to the capital reserve depend on a combination of each individual bank's 
profitability, their willingness to use their profits to expand their 
capital base, as opposed to salary, dividend and bonus payments etc., and any 
changes in regulatory levels on their capital 
that have been imposed. One of the responses to the 2007 credit crisis
for example was a regulatory requirement that banks increase their capital 
ratio by 2018, and this has probably had a contractionary effect on some
economies over the last 5 years. 
\par
However, we do not rule out that in countries that still have reasonably
comprehensive reserve requirements, reserve regulation is not also playing
a role. The history of the reserve requirement since the 1940's is for
a progressive reduction, as computerized handling considerably reduced
the time required for inter-bank processing, and hence 
liquidity\footnote{Except in China where attempts have been made to
actively use the reserve requirement for regulation.} Reductions to
the reserve requirement will allow credit and money supplies to 
expand over time, up to the new limit. If this expansion is simultaneously
throttled by the capital requirements, then the gradual arrival at the new limit
may not be obvious for several years until after it has occurred. 
\section{Conclusion}
Discrete event simulations have been widely used in engineering as
a tool for both exploring and testing regulatory controls on complex system
behaviour. While not all systems lend themselves to this approach,
the banking system is by its very nature 
a discrete event system, and we believe that this approach has
considerable potential in analysing the response of banking
systems to changes in the regulatory and financial framework.
\par
Accounting rules are determined by 
accounting principals which are typically determined by a national
standard boards and regulatory oversight bodies. Many countries are 
converging on the International Financial Reporting Standards (IFRS) established
by the International Accounting Standards Board. These rules are 
primarily designed to provide an accurate view of the business state
of the individual companies that use them, rather than the stability
of the banking system within which they operate. 
Consequently the sensitivity to conditions demonstrated in this paper 
of significant concern, since it indicates that the banking system 
can be influenced by features of accounting and book keeping practices
that are in practical terms invisible to macro-economic analysis. 
\par
Another matter deserving re-examination is the precise influence
of interest rate changes on modern banking systems.
The idea that the system can be controlled through
interest rate manipulations is based on theoretical arguments 
originating from the early 20th
century gold standard system as operated in Britain. There does not
however appear to be any concrete proof that this "control" operates in the
way commonly claimed. 
Nor does there appear to have been any attempt made to verify that it applies
to banking systems besides those of the early 20th century gold standard,
if indeed it was applicable then.
The results in this paper with respect to interest rates 
suggest that these assumptions do not hold true for Basel 
regulated systems, and this is supported by the experience in Iceland
between 2005-2007 where an acceleration in lending led to a doubling
of liability deposits in the Icelandic banking system, despite
the central bank raising the base interest rate to over 18\% to prevent it.
\par
The illusion of control is a well known phenomena in psychology originating
from work by Langer in 1975\cite{langer.1975}. The dominant characteristics
of banking systems: continuously varying rates of expansion in the credit 
and money supply
over long periods, periodic shocks, and
interactions within a complex and dynamic economy, in conjunction with
what appears to be considerable
sensitivity to relatively small changes provides an ideal
setting for this phenomena. Notable events will always be occurring in the
economy somewhere, and without recourse to a detailed analysis based on 
fundamental mechanisms, it is impossible to ascribe causal relationships 
with any certainty. As a consequence, while economic theories and their 
mathematical models can find a source of supporting empirical evidence for 
theories applied to short time scales - longer periods present considerably
more challenge.
\par
We can for example, use the evidence from the simulations presented here 
to suggest an alternative narrative 
to that of the popular "zero-bound" hypothesis. 
If the influence of the banking system is linked to the period of its loans,
rather than short term interest rate changes, then
the banking system's behaviour must evolve over decade long time scales, 
rather than months. In the absence of any direct influence
of interest rates on lending supply, and several indications that
second order effects are directly rather than inversely influenced by
interest rates, it would seem reasonable to conclude that increasing 
interest rates in a Basel regime may increase the rate of money 
and credit expansion, rather than reducing it as generally believed.
\par
We can consequently hypothesize that a long term process 
is occurring, where well intentioned policy interventions are 
having the opposite effect to that desired.  As liquidity issues 
periodically arise in the banking system, either due to inevitably occurring 
imbalances caused by long term interest 
rate flows, or by an excess of lending in the larger 
economy, a higher loan default rate occurs than can be absorbed by the 
system on either the asset or liability side of the balance sheet.  
Initially this manifests itself as a reduction in credit
availability, which then triggers a Fisher cascade failure\cite{fisher.1933}
 owing to the
general reliance within the economy on a continuously expanding supply
of credit and money from the banking system. This reduction in lending 
triggers intervention by the central banks to stimulate the economy by 
dropping interest rates.  Since however the system is supply rather than 
demand constrained, this hinders the ability of 
banks to increase their capital and absorb loan losses, leading to
a further reduction in economic activity.\footnote{This may be partially offset 
if lower interest rates act to reduce loan defaults by lowering the 
repayment burden on the borrower.} Eventually the banking system begins to 
once again expand lending, as loan defaults are finally absorbed, and the 
central bank shortly thereafter, following accepted economic theory, reacts by raising 
interest rates to prevent the economy overheating. In the short term this 
again has the opposite effect to that desired, as higher interest rates 
increase bank income\footnote{Nominally, bank income is interest
rate neutral since banks also pay interest on their savings accounts. 
However not all accounts attract interest, and banks also have more leeway
to increase spreads in a high interest rate period than a low one. It
could be argued this is an example of the zero bound influencing
the system, but again in an opposite direction to that generally believed.}, 
allowing increased capital expansion and lending\footnote{The Icelandic
banking system in the 2005-7 years presents an extreme example of this 
with some highly questionable methods being used to increase capital.}, 
and the central bank in its turn raises interest rates higher.  
Eventually this triggers a 
new credit crisis either through the build up of liquidity imbalances 
caused by long term interest rate flows, or by causing excessive loan 
defaults as borrowers 
are unable to meet their interest repayments. Since the early 1980's this series
of events would also 
have been exacerbated by the widespread use of loan securitization, which
we can infer from the experiment involving cross bank lending, would
act to further unbalance liquidity flows within the 
banking system and place increased stress on the asset side of the system.
\par
Whatever may be thought of this particlar argument, the larger problem remains.
How do we determine if or when, this or any other explanation is correct? 
The banking system presents a complex set
of transactions, interacting within a dynamic economy. 
It is not the case after all that current
economic theory is lacking plausible explanations for observed
phenomena, what is absent is any reliable mechanism for determining
which theory is correct, and given the long history of systemically
effecting changes to the banking system, when.
\par
The approach outlined in this paper offers a way out of this impasse.
Threadneedle is not designed to simulate any particular banking system, 
but rather to allow banking system simulations to be constructed. This allows 
experimental banking systems to be constructed and used to isolate
the behaviour of the different components of the system, accounting
treatments, financial
instruments, or regulatory requirements. These results can then be used
as the basis for larger and more realistic economic 
simulations and models tailored to the regulatory peculiarities of individual 
national banking systems.
\par
More work is needed to develop the features available
within Threadneedle, in particular a richer set of financial
instruments is required, as well as central bank, commercial bank
and borrower behaviours, and the inclusion of international banking,
foreign exchange and stock and commodity markets. This will require 
research into determining
the actual accounting mechanisms being used for these operations, and
any regional or national differences that exist, but lies well
within the scope of current computer technology.
\par
Basing simulations on double entry book keeping operations also opens up 
the eventual possibility of linking economic simulations. 
Double entry book keeping provides
a common set of financial operations identical to those
used in the actual financial system, and as a consequence also provides a 
common reference point that can be used 
to link simulations together. This opens up the possibility of
detailed simulation of the economy at the national and international
level by multiple teams. It should be acknowledged though that constructing 
large complex simulations of poorly understood systems, however accurate 
they may be, is not necessarily a step forward. They are likely 
to be just as difficult to analyze as their real life counterparts. We suspect 
that the real power of
systems like Threadneedle will be to allow small systems to be constructed
that are tractable to analysis, and teaching, which can then be extended
to larger systems, as a more strongly founded understanding of banking
and its interaction with the economy is progressively established.
\bibliography{finance}
\raggedright
\bibliographystyle{unsrt}

\end{document}